\documentclass[12pt]{article}

\usepackage{graphicx}
\providecommand{\U}[1]{\protect\rule{.1in}{.1in}}

\begin{document}

\hyphenation{Min-kow-ski} \hyphenation{cosmo-logical}
\hyphenation{holo-graphy} \hyphenation{super-symmetry} 
\hyphenation{super-symmetric}

\rightline{VPI-IPNAS-08-05}
\centerline{\Large \bf } \vskip0.25cm \centerline{\Large \bf } \vskip0.25cm
\centerline{\Large \bf Towards a spin dual of the fractional quantum Hall effect}
\vskip0.25cm
\vskip 1cm

\renewcommand{\thefootnote}{\fnsymbol{footnote}}
\centerline{{\bf J. J. Heremans\footnote{heremans@vt.edu} and
Djordje Minic\footnote{dminic@vt.edu}}} \vskip .5cm
\centerline{\it Department of Physics, Virginia Tech} \centerline{\it Blacksburg, VA 24061, U.S.A.}

\begin{abstract}
Electromagnetic duality between the Aharonov-Bohm and the Aharonov-Casher
quantum mechanical phases predicts the existence of a new collective state of
matter which can be regarded as a spin dual to the fractional quantum Hall
effect. The state, induced by electric fields, is driven by effective
spin-spin interactions. We derive experimental and materials conditions of
spin-spin interactions and electric fields under which the new state may be observed.
\end{abstract}

\vskip .5cm

\setcounter{footnote}{0} \renewcommand{\thefootnote}{\arabic{footnote}}

\newpage

Electromagnetic fields can influence quantum mechanical systems by generating
quantum phases. The Aharonov-Bohm (AB) effect \cite{1} describes the quantum
phase generated by a magnetic field enclosed by the trajectory of an
electrical charge. A dual to the AB effect exists, whereby an electric field
generates a phase along the trajectory of a magnetic moment. This quantum
phase is referred to as the Aharonov-Casher (AC) phase \cite{2}. The concept
of electromagnetic duality has been fruitfully utilized in modern quantum
field theory to study the otherwise intractable strongly-coupled limit of
various theories, starting from better understood weakly-coupled ones
\cite{3}. The approach has illuminated how apparently disconnected problems
are related as respectively strongly and weakly-coupled limits of the same
core phenomena. In this letter we explore an application of duality to
correlated electron systems and address whether the approach may lead to new
collective states of matter and observable effects, specifically akin to the
fractional quantum Hall effect (FQHE) \cite{4}. We find that the AC phase
carries deep implications for phenomena in correlated electron systems as
likewise the AB phase has proven to significantly impact charge transport in
the solid state. Indeed, signatures of the AB phase abound in the solid state,
leading to phenomena such as oscillatory effects in mesoscopic rings \cite{5},
weak-localization \cite{6}, universal magnetoconductance fluctuations
\cite{7}, the creation of composite particles in the FQHE \cite{8}, and flux
quantization in superconductors.

The role of the AB phase in the FQHE is well known, and hence the study of a
dual to the FQHE can start with an exploration of the dual phase, the AC
phase. The expression below for the AC phase emphasizes the duality with the
AB phase (in SI units, AC phase at left, AB phase at right):
\begin{equation}
\Delta\phi_{AC}=\frac{1}{\hbar c^{2}}\int_{C}\vec{\mu}\cdot(\vec{E}\times
\vec{dl}),\quad\Delta\phi_{AB}=\frac{1}{\hbar}\int_{C}q(\vec{A}\cdot\vec{dl}).
\end{equation}
Here $\Delta\phi_{AC}$ denotes the AC phase, $\Delta\phi_{AB}$ the AB phase,
$\vec{\mu}$ the particle's magnetic moment, $q$ the particle charge, $\vec{E}$
the electric field, $\vec{A}$ the magnetic vector potential, $\vec{dl}$ a line
element of the trajectory, and $c$ the velocity of light. The ring geometries
in Fig.1 help in visualizing the AC and AB effects. For simplicity in the
figure and without impinging on generality, $\vec{\mu}$ is assumed
perpendicular to $\vec{E}$. The expression for AC in Eq.1 results from a
permutation of the expressions obtained in Ref.2, and can formally be obtained
by introducing an effective $q\vec{A_{eff}}=(1/c^{2})\vec{\mu}\times\vec{E}$.
The duality between the AB and AC effects arises from the topological
equivalence of, on one hand, a closed path of a charge $q$ around a local
magnetic flux tube and, on the other hand, a closed path of a local magnetic
flux around a charge $q$ (generating $\vec{E}$) \cite{9}. In neither the AB or
AC effect do the magnetic field $\vec{B}$ or $\vec{E}$ respectively effect a
force \cite{9, 10}. The AC phase implied by Eq.1 is properly a dynamical phase
\cite{11}, but a Berry's phase may additionally arise if $\vec{\mu}$ evolves
over its trajectory \cite{12}.

\begin{figure}
\begin{center}
\includegraphics[width=2in]{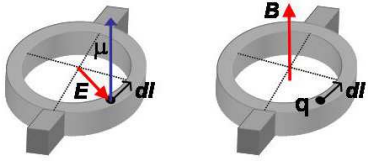}
\caption{Schematic ring geometries to visualize the duality between the
Aharonov-Casher (left) and Aharonov-Bohm (right) effects. Particle paths
follow line elements $\vec{dl}$. For simplicity, the magnetic moment
$\vec{\mu}$ is represented perpendicular to the radial electric field
$\vec{E}$.}%
\end{center}
\end{figure}

Experimentally, the AC effect was observed using neutron beam interferometry
\cite{13}. The duality in the solid state of AB and AC phases was emphasized
by Mathur \cite{14} in a theoretical study of antilocalization. Other
theoretical studies emphasize the role of the AC phase in interference effects
under spin-orbit interaction (SOI) in mesoscopic rings \cite{11, 12}.
Experimental efforts so far perform AB-type experiments on mesoscopic rings in
a variable perpendicularly applied $\vec{B}$, approaching the AC phase as a
modification to the AB phase under strong SOI, rather than as a dual effect
from which new states of matter may arise. Ring arrays in InGaAs
heterostructures were studied \cite{15}, as well as single rings
on HgTe \cite{16}. Current experiments aiming to demonstrate the AC phase in the 
geometry of Fig.1 will help demonstrate the duality implicit in Eq.1. 

The many-body Hamiltonian describing the FQHE collective state of matter
reads:
\begin{equation}
\frac{1}{2m^{\ast}}\sum_{j}[-i\hbar\vec{\nabla_{j}}-q\vec{A}(\vec{r_{j}}%
)]^{2}+\sum_{j<k}V_{C}(\vec{r_{j}}-\vec{r_{k}})
\end{equation}
where $\vec{\nabla_{j}}$ represents the gradient with respect to coordinate
$\vec{r_{j}}$ of the $j^{th}$ electron and $m^{\ast}$ the effective mass. The
pair-wise Coulomb interaction energy $V_{C}$ causes the FQHE. A potential term
expressing a neutralizing background charge has been omitted for simplicity
\cite{17}. Similarly a dual may be constructed by substituting $q\vec{A_{eff}%
}=(1/c^{2})\vec{\mu}\times\vec{E}$ and considering the many-body Hamiltonian
for interacting $\vec{\mu_{j}}$ or spins in an applied $\vec{E}$:
\begin{equation}
\frac{1}{2m^{\ast}}\sum_{j}[-i\hbar\vec{\nabla_{j}}-(1/c^{2})\vec{\mu_{j}%
}\times\vec{E}]^{2}+\sum_{j<k}V_{C}(\vec{r_{j}}-\vec{r_{k}})+\sum_{j<k}%
V_{S}(\vec{\mu_{j}},\vec{\mu_{k}})
\end{equation}
with $V_{S}(\vec{\mu_{j}},\vec{\mu_{k}})$ denoting a spin-spin interaction
energy if $\vec{\mu_{j}}=g^{\ast}\mu_{B}S_{j}$ (here $S_{j}$ is the particle
spin, $g^{\ast}$ is the electron $g$ factor in the material, and $\mu
_{B}=e\hbar/2m_{e}$ is the Bohr magneton). We explore whether dominantly
strong spin-spin interactions $V_{S}$ may in the presence of an applied
$\vec{E}$ result in a dual of FQHE, referred to as the spin dual quantum Hall
effect (SDQHE). Spin-spin interactions have received substantial recent
attention for their role in the pairing mechanisms suggested in
superconductivity \cite{18}, and can in select systems play a dominant role.

As proposed by Laughlin \cite{19} for two-dimensional systems (2DSs), the
ground state wave function for Eq.2 at a FQHE state characterized by odd
fractional Landau level filling factor $\nu=1/(2n+1)$, is:
\begin{equation}
\Psi_{2n+1}(z_{1},...,z_{n})=\prod_{j<k}^{N}(z_{j}-z_{k})^{2n+1}\exp(-\frac
{1}{4l_{B}^{2}}\sum_{i=1}^{N}|z_{i}|^{2})
\end{equation}
where $n$ is an integer, $z_{j}=x_{j}+iy_{j}$ are the complex coordinates of
the $j^{th}$ electron, and $l_{B}=\hbar/eB$ \ is the magnetic length. The
exponential term of the Landau level wave functions is multiplied by the
Jastrow factor $(z_{j}-z_{k})^{2n+1}$ by which three functions are fulfilled
\cite{20}: 1) the many-body wave function acquires the correct antisymmetry
under particle exchange, and 2) $\vec{B}$ is accounted for with the correct
$\nu$, and 3) the correlations due to Coulomb interaction are satisfied. The
Jastrow factor attaches to each electron a flux tube of $2n+1$ magnetic
Aharonov-Bohm flux quanta (one flux quantum is $\phi_{0}=h/e$), ensuring an
overall correct $\nu$ for the homogeneously applied $\vec{B}$. The addition of
magnetic flux quanta also decreases the Coulomb interaction energy since the
Jastrow factor reduces the wave function amplitude if electrons approach each
other. Functions (2) and (3) of the Jastrow factor in Laughlin's wave function
embody the core of the FQHE, and are also amenable to the SDQHE. We envision a
dual state where an interacting spin system under an applied $\vec{E}$ is
described by a Laughlin-type wave function. To build the SDQHE, the magnetic
flux quantum in the FQHE is substituted by an electrical line integral
quantum, $Y_{0}$ defined below (see also Eq.1 and Fig.1) and we assume a
spin-spin interaction $V_{S}$ dominant over the Coulombic $V_{C}$.
Experimental systems for the realization of this crucial condition are
discussed in later sections. For the emergence of the FQHE, the exact form of
$V_{C}$ is not important, as it suffices that the Jastrow factor captures the
major part of the interactions, leaving residuals unable to change the ground
state. Likewise, we expect latitude in the exact form of $V_{S}$ causing the
SDQHE. The dual described by the corresponding Laughlin wave function is a new
correlated Fermionic state, as is understood from the odd exponent in the
terms $(z_{j}-z_{k})^{2n+1}$. A description of the FQHE has evolved whereby
the FQHE is understood as the integer quantum Hall effect (IQHE) of composite
Fermions \cite{20}. Particles similar to the composite Fermions are expected
to have significance in the SDQHE, and will likewise possess Fermionic
statistics. A spin dual to the IQHE is briefly discussed below.

To establish similarities to quantum Hall geometries we consider a 2DS in the
x-y plane. In quantum Hall geometries, $\vec{B}=(0,0,B)$ may be generated from
the symmetric gauge potential $\vec{A}=(-By/2,Bx/2,0)$. In a gedanken
experiment for the SDQHE, we assume the spins $\vec{\mu}$ are aligned parallel
to $z$, resulting from ferromagnetic interactions. From $q\vec{A_{eff}%
}=(1/c^{2})\vec{\mu}\times\vec{E}$ a symmetric gauge linear in coordinates
then signifies a radially increasing in-plane $\vec{E}$, as encountered in a
uniformly charged cylinder. In this gedanken experiment, wave functions
\cite{21}, energy levels and degeneracies of a magnetic moment in $\vec{E}$
faithfully mimic a charge in $\vec{B}$. For other profiles of $\vec{E}$,
energy levels may not be equally spaced and level degeneracies may differ,
without impact, however, on the development of the SDQHE. Without impact on
generality while allowing a more transparent treatment, we will below consider
the special case of spins oriented along $z$. Beyond this special case, the
consequences of the full $SU(2)$ spin symmetry as well as antiferromagnetic
alignment promise a rich spectrum of behavior in the SDQHE. For $\vec{\mu}$
parallel to $z$, the above similarity ensures that akin to the IQHE, broken
translational invariance at sample edges will result in edge states. Arguments
of gauge invariance and charge conservation \cite{22} then lead to a spin dual
of the IQHE. Since the carrier spin is associated with a charge, electrical
transport characteristics akin to the IQHE and FQHE are expected for the
corresponding spin duals. The energy level structure is carried by spin,
whereas charge conservation preserves the quantization argument. In this broad
context, routes distinct from dualization, involving SOI and the role of
$SU(2)$ in the quantum Hall effect have been discussed previously \cite{23}.

From the geometries in Fig.1 and the argument involving the radial in-plane
$\vec{E}$ above, we may derive general experimental geometries in which the
SDQHE may be observed. $\vec{B}$ perpendicular to the plane of the 2DS
introduces the flux quanta $h/e$ in the FQHE. The SDQHE will require $\vec{E}$
in the plane of the 2DS, which introduces an electrical line integral quantum,
$Y_{0}$. In analogy with the flux periodicity $\phi_{0}$ observed in
mesoscopic rings due to the AB phase, an AC periodicity can be expressed by
stating:
\begin{equation}
\frac{1}{\hbar c^{2}}\int_{C}\vec{\mu}\cdot(\vec{E}\times\vec{dl})=2\pi p
\end{equation}
with $p$ an integer. With the projection of $\vec{\mu}$ perpendicular to the
line integral equated to $\mu_{B}$, the periodicity in $\vec{E}$ is deduced
from:
\begin{equation}
Y_{0}=\int_{C}\vec{E}\times\vec{dl}=\frac{4\pi m_{e}c^{2}}{e}=6.42\times
10^{6}V
\end{equation}
where $Y_{0}$ carries in the AC phase the same role as $\phi_{0}$ carries in
the AB phase. Other authors have derived quantization conditions for the AC
phase, and have expressed the phase as an addition to the AB phase \cite{14,
24}. In vacuum, the AC phase for an electron spin remains small for
technically achievable magnitudes of $\vec{E}$. Even for a macroscopic sample
where the line integral courses over a circumference of order 1 \textit{mm}
the vacuum value for $Y_{0}$ predicts that very substantial $E\sim10^{10}V/m$
are necessary, higher than the breakdown field of many electronic materials
($\sim10^{7}V/m$). In materials with high SOI, however, the required $E$ is
much reduced \cite{14, 24}, providing an avenue for experimental observation
of the SDQHE. By considering a small section of material wherein the radial
field lines appear approximately parallel (a conformal mapping argument), we
conclude that a Hall bar geometry with an in-plane $\vec{E}$ perpendicular to
the current direction can support the SDQHE, as illustrated in Fig.2. Under
dominant $V_{S}$, spin and charge both will interact with the applied $\vec
{E}$, and experiments should be designed to allow the data to distinguish the
interactions. Particularly in semiconductors, experiments should also account
for gating and the electrostatic AB effect (hitherto not observed) \cite{25}.

\begin{figure}
\begin{center}
\includegraphics[width=3in]{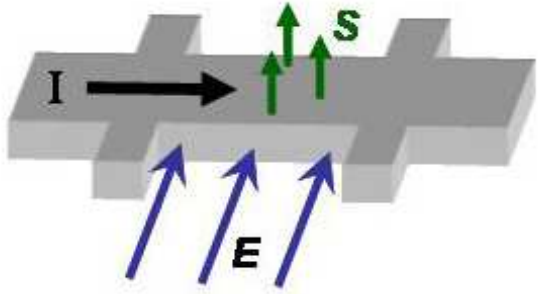}
\caption{Hall bar geometry with in-plane electric field $\vec{E}$,
perpendicular to current direction (\textbf{I}). We consider only the special
case of spins (\textit{S}) oriented perpendicularly to $\vec{E}$.}%
\end{center}
\end{figure}

The FQHE appears in 2DSs where the ratio of the average Coulomb interaction
energy $<U_{C}>$ to kinetic energy is large. The parameter $r_{C}%
=<U_{C}>/E_{F}$, where $E_{F}$ is the Fermi energy, expresses this ratio. In
GaAs/AlGaAs 2DSs, $r_{C}\sim2...30$, high values allowing the observation of
the FQHE. The parameter $r_{C}$ is for the SDQHE recast as $r_{S}%
=<U_{S}>/E_{F}$ with $<U_{S}>$ the average spin-spin interaction energy. For
comparison we first consider magnetic dipole-dipole interactions with parallel
spin alignment (achieved by a weak constant $\vec{B}$ parallel to $z$ with $B$
much below the FQHE regime). With values for $m^{\ast}$ and $g^{\ast}$ typical
of semiconductors, we find $r_{S}\sim10^{-6}$, a small ratio expected from the
higher-order dipole interactions. It is hence advisable to identify systems
with stronger spin-spin interactions. Magnetic exchange interactions between
itinerant electron or hole spins form promising candidates. The carrier
systems of interest should also allow for the application of $\vec{E}$. Dilute
magnetic semiconductors, particularly Mn-doped III-V semiconductors, may unite
these properties. Exchange between the Mn spins and the hole spins lead to
effective large hole spin-spin interactions of the form $V(\vec{\mu_{j}}%
,\vec{\mu_{k}})=J_{jk}(\vec{r})\vec{\mu_{j}}\cdot\vec{\mu_{k}}/(g^{\ast}%
\mu_{B})^{2}$ . The exchange mechanism, still under discussion, determines the
spatial dependence of $J_{jk}(\vec{r})$ [26]. An estimate can be obtained from
$r_{S}\sim k_{B}T_{C}/E_{F}$ with $T_{C}$ the Curie temperature and $k_{B}$
the Boltzmann constant. For representative hole densities in Mn-doped III-V
semiconductors ($3\times10^{26}m^{-3}$) and $T_{C}\sim150K$, we derive
$r_{S}\sim0.1$. While still lower than $r_{C}$, this value in 3-dimensional
materials heartens experimental efforts. Yet, an experimental complication in
dilute magnetic semiconductors may reside in the high concentration
($\sim1.5\%$) of Mn, resulting in substantial disorder likely deleterious to
the SDQHE as it is for the FQHE. In fact, band transport in GaMnAs is
questioned \cite{27}. In other systems, at the border of ferromagnetic or
antiferromagnetic transitions spin susceptibilities increase and strengthen
induced spin-spin interactions. Recent insights in unconventional
superconductivity \cite{18} suggest that these increased spin-spin interaction
can overcome Coulomb repulsion.

In conclusion duality between the AB and AC phases, when applied to the FQHE,
predicts a novel correlated state of matter. The state, a spin dual to the
FQHE with signatures akin to the FQHE, may appear under electric fields in new
materials with strong spin-spin interactions.

\vskip .5cm

\textbf{{\Large Acknowledgements}}

\vskip .5cm

We thank R. L. Kallaher and K. Park for helpful discussions. {\small JJH} is
supported in part by the National Science Foundation under contract
DMR-0618235. {\small DM} is supported in part by the U.S. Department of Energy
under contract DE-FG05-92ER40677.

\end{document}